\documentclass[preprint,journal]{vgtc}            

\onlineid{0}

\vgtccategory{Research}

\vgtcpapertype{please specify}

\newcommand{\brian}[1]{\textcolor{orange}{}}

\include{header}

\newcommand{\toolname}{ZipLine\xspace}

\usepackage{xcolor}
\newcommand{\revised}[1]{{#1}}

\title{\toolname: Visual Analysis of Multivariate Graphs with Predicate Logic}

\author{%
  \authororcid{Sjoerd Vink}{ 0009-0006-1094-3725},
  \authororcid{Suyang Li}{0009-0000-6375-182X}, 
  \authororcid{Brian Montambault}{0009-0000-1988-406X}, 
  \authororcid{Michael Behrisch}{ 0000-0002-1102-103X},
  \authororcid{Mingwei Li}{0000-0002-0457-8091}, 
  and  \authororcid{Remco Chang}{0000-0002-6484-6430}
}

\authorfooter{
  \item
  	Sjoerd Vink is with Utrecht University and Tufts University.
  	E-mail: s.a.vink@uu.nl
      \item
  	Suyang Li, Brian Montambault, Mingwei Li, and Remco Chang are with Tufts University.
  	E-mail: \{suyang.li, brian.montambault, mingwei.li, remco.chang\}@tufts.edu
      \item
  	Michael Behrisch is with Utrecht University.
  	E-mail: m.behrisch@uu.nl
}

\abstract{
    Multivariate graphs unite two distinct data perspectives: a topological structure defined by nodes and edges, and attribute data associated with each node.
    Analyzing such graphs therefore requires reasoning across two complementary spaces.
    However, existing systems typically emphasize the analysis of one space at a time, focusing either on topology or on attributes.
    As a result, exploration, analysis, and pattern discovery that depend on their interaction remain difficult.
    In this paper, we present \toolname, a system designed to support integrative analysis of multivariate graphs by bridging both topology and attribute spaces.
    \toolname introduces a predicate language that enables analysts to express patterns involving topology, node attributes, and neighborhood relations with a unified formalism.
    The system further provides a predicate-learning algorithm that maps analyst interactions across both topology (e.g., subgraph selection) and attribute views (e.g., value brushing), into the predicate language, enabling learned expressions that bridge the two spaces.
    This integrative approach supports iterative analysis by enabling analysts to refine patterns through coordinated reasoning over topology and attributes.
    We demonstrate \toolname through three case studies in energy infrastructure, cybersecurity, and drug discovery analysis.
    The results show that \toolname enables expressive multivariate graph analysis through unified reasoning across topology and attributes.
}


\teaser{
  \centering
  \includegraphics[width=0.9\linewidth]{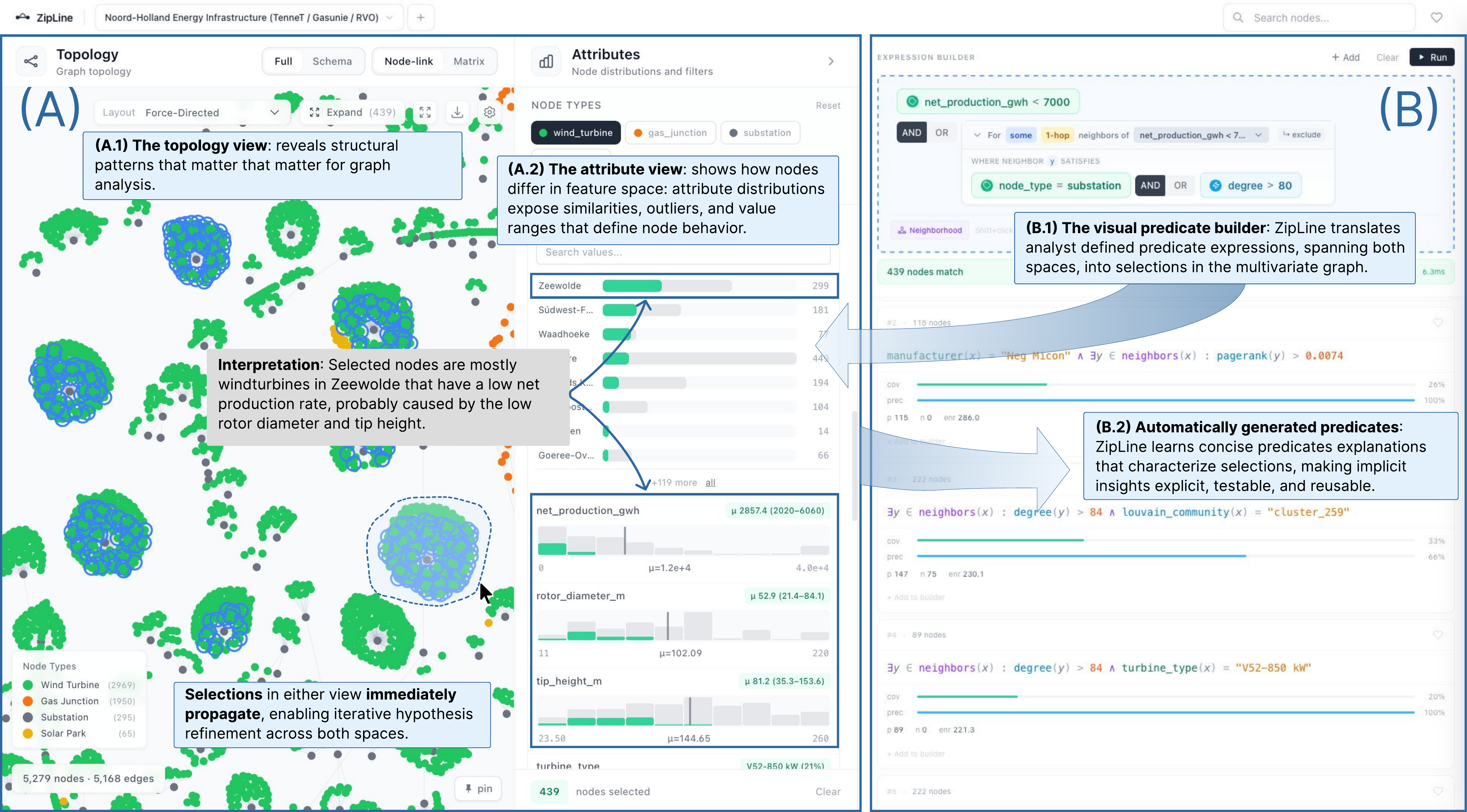}
    \caption{\textbf{Overview of the \toolname interface.}
    \textbf{(A)} \toolname enables joint analysis of topology and attributes through two coordinated views:
    \textbf{(A.1)} the topology view shows connectivity, clusters, and paths;
    \textbf{(A.2)} the attribute view shows distributions and value ranges.
    Selections originate in either view and propagate across both, enabling iterative refinement of cross-space patterns.
    \textbf{(B)} \toolname connects interactions to formal representation through predicates.
    \textbf{(B.1)} Analysts construct expressions that jointly filter topology, attributes, and neighborhoods to define precise graph selections.
    \textbf{(B.2)} The system derives predicates from selections, producing concise descriptions of how both spaces jointly distinguish them.
    }
    \vspace{-0.5em}
  
  \label{fig:system}
}

\graphicspath{{figs/}{figures/}{pictures/}{images/}{./}}

\usepackage{tabu}
\usepackage{booktabs}
\usepackage{lipsum}
\usepackage{mwe}
\usepackage{tabularx}
\usepackage{mathptmx}
\usepackage{amsmath}
\usepackage{amssymb}
\usepackage{mathtools}

\usepackage{tikz}
\newcommand\submittedtext{%
  \footnotesize
  This work has been submitted to the IEEE for possible publication.
  Copyright may be transferred without notice, after which this version
  may no longer be accessible.%
}
\newcommand\submittednotice{%
  \begin{tikzpicture}[remember picture,overlay]
    \node[anchor=south,yshift=10pt] at (current page.south)
    {\fbox{\parbox{\dimexpr0.65\textwidth-\fboxsep-\fboxrule\relax}{%
      \submittedtext}}};
  \end{tikzpicture}%
}

\begin{document}

\maketitle
\submittednotice

\section{Introduction}

A multivariate graph represents entities as nodes connected through edges, where each node is associated with multiple attributes describing its properties.
By combining structural connectivity with attribute data, multivariate graphs provide a natural representation for complex systems and are widely used in domains such as cybersecurity\cite{abu2021domain}, biomedicine\cite{chandak2023building}, and supply chain analysis\cite{mageto2021big}.
For example, in a cybersecurity network, nodes may represent devices with attributes such as the operating system they run or a vulnerability score, while edges capture communication between devices.
Analytical questions in such settings often arise from the interaction between structure and attributes, requiring analysts to reason jointly over both spaces\cite{wattenberg2006visual}.

Despite this expressiveness, the dual nature of multivariate graphs introduces challenges for analysis.
First, when patterns span both spaces, analysts must translate findings between structural and attribute representations.
This becomes difficult when visualization tools emphasize one representation or lack mechanisms for integrated analysis.
Second, reasoning over structure and attributes relies on different cognitive processes\cite{amar2005low, lee2006task,williams2023data}.
Tabular analysis emphasizes value-based operations such as retrieval and aggregation, whereas graph analysis requires relational reasoning over nodes, edges, and paths.
Without explicit support to bridge these representations, the cognitive burden on the analyst remains high.

\revised{In this paper, we present a technique for formalizing and inducing patterns in multivariate graphs, instantiated in the \toolname visual analytics system.}
\revised{The central contribution is a predicate formalism that provides a common representation for reasoning across topology and attribute spaces.}
\revised{The formalism defines a predicate language for describing and selecting patterns involving structural properties, node attributes, and neighborhood relations.}
Node predicates capture attribute and structural filters, while bounded neighborhood predicates express quantified filters over finite neighborhoods, such as requiring a minimum number or proportion of adjacent nodes satisfying a predicate.
The language excludes unbounded path traversal and recursion, ensuring deterministic evaluation and a finite search space. 

Building on this formalism, \toolname provides a predicate learning algorithm that induces logical expressions from analyst-defined selections.
Given nodes or data identified during exploration, the algorithm searches for predicates that characterize the selection by combining attribute filters, structural properties of the corresponding subgraph, and criteria on their neighborhoods (e.g., requiring that a node has at least $k$ neighbors).
For example, analysts may select nodes in the topology view (e.g., a node-link diagram) or brush over value ranges in the attribute view (e.g., a histogram barchart).
In both cases, the algorithm generates predicates that integrate attribute and topological information to represent the selected nodes or values across both spaces.
This process externalizes implicit visual observations as explicit logical expressions that can be inspected, refined, and reused.
Because predicates operate directly on the data representation, the approach is independent of visualization design.

We evaluate \toolname through three domain use cases in energy systems, cybersecurity, and structural biology using real-world multivariate graphs.
In energy systems, we identify low-output turbines that occupy structurally critical positions in the transmission backbone, showing assets whose failure would disproportionately impact grid stability\cite{tennet_nh_energy}.
In cybersecurity, we analyze the MITRE ATT\&CK graph\cite{mitre_attack_stix_data} to distinguish tactical patterns of threat actors, showing how differences in technique connectivity and usage reflect fundamentally different operational strategies.
In structural biology, we analyze drug repurposing in a biomedical graph\cite{chandak2023building} to identify molecular bridge mechanisms that explain off-label drug effects, identifying intermediate proteins that connect disease contexts through shared regulatory pathways.
Across these use cases, we show that iterating between interactive predicate construction and automated induction enables the identification and explanation of patterns, and that selected subgraphs can be translated into logical expressions that support reasoning across domains.

\section{Related Work}

Our work builds on prior research in multivariate graph visualization, graph querying, and predicate-based rule induction.

\subsection{Approaches to Analyzing Multivariate Graphs}

\vspace{0.5em}\noindent\textbf{Multivariate Graph Visualization.}
A large body of work focuses on visually integrating topology and attribute information in multivariate graphs.
Node attributes are commonly encoded using visual channels such as size, color, position, or shape\cite{borisjuk2005integrating, frasincar2006adapting, perer2006balancing}, and have also been incorporated directly into layout computation for node-link diagrams\cite{archambault2008grouseflocks, fekete2003overlaying}. 
Building on these ideas, PivotGraphs\cite{wattenberg2006visual} present an aggregated network representation in which nodes are arranged within a grid defined by two categorical attributes, enabling the exploration of structural patterns conditioned on attribute combinations. 
GraphTrail\cite{dunne2012graphtrail} emphasizes interaction by capturing and organizing user actions into an explicit interaction history that supports reflection and revisitation during multivariate network exploration.
Similarly, Van den Elzen and Van Wijk\cite{van2014multivariate} introduce a detail-to-overview approach based on interactive selection and aggregation.
VIGOR\cite{pienta2017vigor} emphasizes exploration and summarization of graph query results using coordinated views.
OnionGraph\cite{shi2020oniongraph} supports hierarchical abstraction where aggregation can be driven by topology, attributes, or both.

These approaches provide rich mechanisms for visually integrating topology and attributes.
In contrast, we connect visualization to formal representation, turning discovered patterns into explicit and reusable specifications.
This brief overview represents only a small subset of the extensive body of work on multivariate graph analysis.
For more information, we refer readers to existing survey papers that systematically cover design spaces in this area\cite{kerren2014introduction, nobre2019state, kale2023state}.


\vspace{0.5em}\noindent\textbf{Graph Exploration and Querying Systems.}
Another line of work investigates systems that support querying and exploration of graph-structured data.
Vistorian\cite{molinero2017understanding} provides visualization techniques tailored to historical and dynamic graphs.
Cytoscape\cite{shannon2003cytoscape} is a widely adopted platform in the life sciences for network visualization and analysis, offering extensive support for attribute inspection and structural metrics.
General-purpose tools such as Gephi\cite{gephi} and LinkQ\cite{li2024linkq} further facilitate graph exploration through coordinated views, interaction history, and query-oriented interfaces.
\revised{These systems primarily support exploration through visual encodings, interaction, and querying.}

Prior work has also explored visual graph querying and how interactive interfaces can support the construction and execution of graph queries through direct manipulation. 
VISAGE\cite{pienta2016visage} enables incremental query construction by allowing users to interactively select nodes, edges, and attributes within a graph visualization. 
Envisage\cite{wen2025envisage} extends this approach by supporting more expressive query constructs and compositional mechanisms for building complex graph queries visually.
Konduit VQB\cite{ambrus2010konduit} focuses on semantic data and provides a visual interface for constructing SPARQL queries, translating graphical query representations into executable query expressions.
ComBiNet\cite{pister2023combinet} supports visual querying and comparison of bipartite, multivariate, and dynamic social networks through coordinated views and interactive filtering.
Pathfinder\cite{partl2016pathfinder} emphasizes the visual analysis of paths in graphs, enabling users to explore and compare alternative connectivity structures.
Graphite\cite{chau2008graphite} introduces a visual query system for large graphs that allows users to specify structural query patterns through interactive graph manipulation.
KGNav\cite{wang2023kgnav} provides a visual query interface tightly integrated with schema-aware navigation and result presentation for knowledge graphs.
\revised{More recent systems such as GraphQ IR\cite{nie2022graphq} and HiRegEx\cite{li2024hiregex} use intermediate representations and grammars to bridge natural language or hierarchical data with formal queries.}

\revised{Related earlier work includes Doleisch et al.\cite{doleisch2003interactive}, who proposed a feature definition language for interactive predicate specification in visualization.}
\revised{Xiao et al.\cite{xiao2006enhancing} integrate first-order logic into visual analysis, enabling analysts to formalize and reuse visual findings as predicates.
ZipLine builds on this principle but differs in scope and mechanism.
It defines predicates over node attributes, derived topological properties, and bounded neighborhoods in multivariate graphs, and learns these predicates from analyst selections.
Predicate explanation systems reviewed below instead derive rules for tabular data, projection patterns, or model behavior.
Our contribution is a predicate language for multivariate graphs and an induction algorithm that learns expressions in this language from analyst selections.}

\subsection{Predicate and Rule Induction}


\vspace{0.5em}\noindent\textbf{Predicates for Data Analysis.}
Predicates are logical filter that define subsets of data.
Such representations are widely used to describe patterns in data in an interpretable and reusable form.
Subgroup discovery\cite{klosgen1996explora, wrobel1997algorithm, atzmueller2015subgroup, herrera2011overview}
focuses on identifying statistically interesting subsets characterized by descriptive predicates.
Similarly, rule-based learning approaches represent patterns as logical expressions composed of attribute-value filter\cite{quinlan2014c4, cohen1999simple, wu2013scorpion}.
Association rule mining\cite{agrawal1993mining} further demonstrates how relationships between variables can be expressed through predicates.
These approaches share the objective of describing data through explicit logical filters that are human-interpretable and directly actionable.
Such predicate-based representations form the foundation for explaining and characterizing patterns in data.

\vspace{0.5em}\noindent\textbf{Predicate Learning and Rule Induction.}
Induction is the process of learning general rules from specific examples.
Generally there are two categories of induction.
Predictive induction focuses on learning from existing data so it can predict the class or outcome of new, unseen examples.
In contrast, descriptive induction looks for meaningful and interesting patterns in data to better understand its structure, without necessarily aiming to make predictions\cite{novak2009supervised}.
Algorithms are available that use exhaustive search strategy to find subgroups.
Search space can be constrained with techniques like anti-monotone property\cite{han2007frequent} or optimistic estimate function\cite{von2009tight}.
An alternative is a beam search\cite{russell1995modern} strategy to optimize the approach.
\revised{We use beam search because neighborhoods greatly expand the search space.}

There are also systems for rule or predicate explanations.
Divisi\cite{sivaraman2025divisi} is an interactive notebook-based tool using a fast approximate subgroup discovery algorithm.
SURE\cite{yuan2022visual} is a visual analytics system that integrates a hierarchical surrogate rules algorithm and interactive surrogate rule visualizations.
DimBridge\cite{montambault2024dimbridge} does predicate induction on user selections in projections.
RuleMatrix\cite{ming2018rulematrix} is an interactive visualization technique to help users with little expertise in machine learning to understand, explore and validate predictive models.
FairVis\cite{cabrera2019fairvis} is a mixed-initiative system that integrates a subgroup discovery technique for users to audit the fairness of machine learning models.
As far as we know, there is no available work on this for multivariate graphs.

\begin{table*}[ht]
    \centering
    \small
    \begin{tabularx}{\linewidth}{p{2.6cm} p{6cm} p{6cm} p{2.0cm}}
    \toprule
    \textbf{Topological Property} & \textbf{Description} & \revised{\textbf{Example Analytical Use}} & \textbf{Type} \\
    \midrule
    Degree & Number of directly connected neighbors & Identifying high-load substation hubs in power grids & Numerical $\in \mathbb{N}$ \\
    Betweenness centrality & Proportion of shortest paths traversing the node & Detecting grid bottlenecks whose failure isolates regions & Numerical $\in [0,1]$ \\
    Closeness centrality & Reciprocal of the mean shortest-path distance & Finding proteins that rapidly reach all disease targets & Numerical $\in (0,1]$ \\
    PageRank\cite{page1999pagerank} & Node importance based on recursive influence from neighbors & Ranking cyber-security threat actors by operational influence & Numerical $\in [0,1]$ \\
    Clustering coefficient & Local edge density within a node's neighborhood & Spotting dense regulatory protein modules & Numerical $\in [0,1]$ \\
    \revised{k-core number\cite{seidman1983network}} & Maximum $k$ such that the node belongs to a $k$-core subgraph & Locating backbone nodes in communication networks & Numerical $\in \mathbb{N}$ \\
    Community (Louvain\cite{blondel2008fast}) & Identifier of the community assigned to the node & Grouping threat actors by shared tactical patterns & Categorical \\
    Component ID & Identifier of the connected component containing the node & Isolating disconnected subgraphs in fragmented networks & Categorical \\
    \bottomrule
    \end{tabularx}
    \vspace{-1em}
    \caption{Node-level topological features capture distinct structural roles and map to domain-specific analytical scenarios as described in Section~\ref{chap:evaluation}.}
    \label{tab:topology_features}
    \vspace{-2em}
\end{table*}
\section{Representational Spaces in Graph Analytics}
\label{chap:reasoning_levels}

A multivariate graph represents relational structure (topology) together with information attached to nodes (attributes).
\revised{Topological properties (see \autoref{tab:topology_features}) can be used to satisfy concrete analytical tasks. For instance, degree identifies critical substation hubs whose failure disproportionately impacts energy grid stability, betweenness centrality exposes bottleneck proteins bridging disease pathways, and community membership groups cyber-security threat actors by shared tactical patterns.}
Edges may carry a type, but do not carry additional attributes.
\revised{Supporting edge attributes would require extending our predicate language with edge-level filters and adapting the induction algorithm accordingly, substantially increasing the complexity of both; we leave this for future work.}
Additionally, we assume a static, simple graph without multi-edges (i.e., not multigraphs).
The formulation applies to both directed and undirected graphs.

Formally, we define a multivariate graph as
$G = (V, E, A, R)$, where:
\begin{itemize}
    \item $V = \{v_1, v_2, \ldots, v_n\}$ is the set of nodes
    \item $R$ is a finite set of edge types
    \item $E \subseteq V \times R \times V$ is the set of typed edges
    \item $A : V \rightarrow \mathbb{D}$ is an attribute function that maps each node to a finite set of attribute key-value pairs
\end{itemize}

Here, $\mathbb{D}$ denotes the attribute domain, i.e., the space of possible attribute records, where each record consists of named attributes with numerical, categorical, or boolean values.

Graph analytics operates across two connected representational spaces: the \emph{attribute space} and the \emph{topology space}.
The attribute space corresponds to the domain $\mathbb{D}$ of node attributes.
\revised{Attributes describe characteristics of nodes, defined separately from the connectivity between them.}
For example, in a protein interaction network, attributes may include category (e.g., protein, enzyme), molecular weight, or activity.
The topology space corresponds to the graph structure $(V, E)$ and captures the connectivity between nodes.
It defines which nodes are connected and thereby also determines the \emph{neighborhood}.
For a node $x \in V$, its (one-hop) neighborhood \revised{$N_1(x)$} is the set of nodes adjacent to $x$.
Each node $y \in N_1(x)$ 
is itself a node in $V$ and therefore has its own attributes and connections in the graph.
From this connectivity, structural properties can be derived, such as degree, centrality measures, and clustering coefficients.
Unlike attributes, these properties are not stored explicitly but are computed from the graph structure.
All topological properties considered in this work (see \autoref{tab:topology_features}) are represented as node-level features derived from the graph structure.

In practice, analytical questions vary in how they relate to these spaces.
Task taxonomies in graph visualization distinguish between attribute-based operations (e.g., retrieving or comparing node properties) and topology-based operations (e.g., identifying connectivity, paths, or structural roles)\cite{lee2006task, amar2005low}.
Some tasks are naturally confined to a single space, such as identifying nodes with extreme attribute values or detecting highly connected hubs.
However, many tasks require reasoning across both spaces, relating structural roles to attribute characteristics.
For example, identifying low-output turbines embedded in structurally critical regions or distinguishing threat techniques based on both their tactical category and their connectivity cannot be answered using a single space.
This motivates the need for approaches that support integrated reasoning across topology and attributes.

\section{A Predicate Language for Multivariate Graphs} \label{chap:predicates}

We introduce a predicate language grounded in first-order logic for defining filters over multivariate graphs.
These filters specify selections of nodes based on attribute, topology, and neighborhoods.
The language provides a mechanism for expressing filters as explicit logical expressions.
It serves as the foundation for the predicate induction algorithm introduced in Section \ref{sec:predicate-learning}.

\subsection{Background: First-Order Logic}

First-order logic (FOL) provides a formal language for expressing filters over entities through predicates and quantifiers.
A predicate is a Boolean-valued function that evaluates properties of entities in a domain.
In our setting, attribute-based predicates are expressed as comparisons over attribute functions.
For example, we model categorical attributes as functions such as $role : V \rightarrow D_{cat}$, and write predicates of the form $role(x) = \text{``protein''}$, where $x$ denotes an entity in the domain. 
Similarly, $molecular\_weight(x) > 500$ asserts that $x$ has a molecular weight greater than 500.
We refer to each atomic condition, such as $role(x) = \text{``protein''}$ or $molecular\_weight(x) > 500$, as a \emph{clause}.
A predicate consists of one or more such clauses.
Clauses can be combined using conjunctions ($\land$) and disjunctions ($\lor$).
This allows predicates to express compound filters such as $(role(x) = \text{``protein''} \land active(x) = true) \lor role(x) = \text{``enzyme''}$.
Logical expressions may be grouped using parentheses to explicitly control evaluation order and scope.
For the purpose of predicate induction that will be introduced in Section \ref{sec:predicate-learning}, we restrict the logical connectives to positive clauses and exclude explicit negation, as negation substantially increases the search space and reduces learning tractability.

Quantifiers extend predicates to express filters over sets of entities.
There are two quantifiers that FOL offers: the universal quantifier ($\forall$) and the existential quantifier ($\exists$).
The universal quantifier asserts that a predicate holds for all entities in the domain under consideration.
For example, $\forall x: role(x) = \text{``protein''} \land active(x) = true$ means that every node in the domain is an active protein.
The existential quantifier asserts that there exists at least one entity in the domain for which the predicate holds.
For example, $\exists x: role(x) = \text{``protein''} \land molecular\_weight(x) > 500$ \revised{asserts that at least one protein has a weight greater than 500.}

\subsection{Attribute and Topology Spaces}

\revised{Building on this first-order logic (FOL), predicates can be used to express filters that define selections within each space described in Section \ref{chap:reasoning_levels}. These predicates operate independently within a single space, relying solely on either attribute-based or topology-based properties without referencing the other space.}
For example, a predicate such as $role(x) = \text{``protein''}$ filters entities based solely on attribute information, while a predicate such as $degree(x) > 5$ filters entities based solely on structural properties derived from the topology.
These predicates describe isolated filters within a single space and serve as the basic building blocks for more complex expressions.

In practice, many graph analysis questions span both \revised{representational} spaces rather than remaining confined to a single one.
The predicate formalism allows filters to explicitly combine attribute-based and topological properties within a single expression.
We refer to such expressions as \textit{cross-space predicates}, which simultaneously reference properties derived from node attributes and from graph structure.
Cross-space predicates are formed by composing single-space predicates using logical connectives.
For example, the expression $degree(x) > 5 \land role(x) = \text{``protein''}$ combines a structural filter with an attribute-based filter to describe nodes that satisfy both filters.
By representing these relationships as predicates, cross-space predicates make the interplay between topology and attributes explicit.

\subsection{Neighborhood and Cardinality Operators}

In addition to structural properties defined on individual nodes, the topology space provides access to neighborhoods, i.e., sets of nodes adjacent to a given node.
To express neighborhoods beyond immediate adjacency, we introduce a neighborhood operator parameterized by a sequence of edge types.
\revised{Let $\Sigma = (r_1, \dots, r_m)$ be a finite sequence of edge types, where each $r_i$ is drawn from the set of edge types $R$ defined by the graph schema; that is, $r_i \in R$ for all $i$.}
The untyped neighborhood $N_k(x)$ collects all nodes reachable from $x$ within at most $k$ hops regardless of edge type, with $k$ specified as an integer subscript.
\revised{Unlike the typed neighborhood, which follows an edge-type sequence exactly, $N_k(x)$ includes nodes at any distance up to and including $k$.}
The immediate neighborhood $N_1(x)$, also written as $\mathit{neighbors}(x)$, is the standard one-hop case.
Neighborhood operators allow predicates to be evaluated over structurally defined sets of nodes.
When combined with quantifiers, they express filters on local graph context.
\revised{For example, $\forall y \in neighbors(x) : active(y) = true$ selects all entities ($y$) directly connected to $x$ whose "active" attribute is true.}

Many graph patterns depend not only on whether neighbors satisfy a predicate, but on how many do.
We therefore allow predicates to filter the cardinality of such structurally defined subsets.
Given a predicate
over neighbors of $x$, we write $|\{\, y \in neighbors(x) \mid P(y) \,\}| \;\bowtie\; k$, where $\bowtie \in \{=, \geq, \leq\}$, to specify that exactly, at least, or at most $k$ neighbors of $x$ satisfy \revised{a predicate} $P$.
For example, $|\{\, y \in neighbors(x) \mid active(y) = true \,\}| \geq 3$ selects entities connected to at least three active neighbors.

\subsection{Result Structure Semantics}

Predicate expressions are evaluated over the graph $G = (V, E, A, R)$ and return sets of variable bindings that satisfy the predicate.
In the simplest case, a predicate $P(x)$ defines a subset of nodes: $\{\, x \in V \mid P(x) \,\}$.
That is, evaluation yields all nodes $x$ for which the predicate holds.

Variables introduced within predicates, for example through neighborhood quantification, are used only to evaluate the condition.
They do not appear in the result unless they are explicitly part of the output expression.
For example, in
$\exists y \in \mathit{neighbors}(x) : role(y) = \text{``enzyme''}$
the variable $y$ is used to test the condition, but the result consists only of nodes $x$ that satisfy it.
When multiple variables are explicitly included, the result consists of tuples of nodes.
For example,
$\{\, (x,y) \mid y \in \mathit{neighbors}(x) \land role(y) = \text{``enzyme''} \,\}$
returns pairs of nodes $(x,y)$ that satisfy the predicate, making the relational structure explicit.
In all cases, predicate evaluation returns a set of nodes or tuples of nodes, depending on which variables are specified.
Edges are not returned explicitly.
However, relations such as adjacency are reflected implicitly through the variable bindings.

The semantics of predicate evaluation are deterministic.
Given a graph and a predicate expression, the resulting set of variable bindings is uniquely determined by the predicate.
This allows predicate expressions to serve as precise and reproducible specifications of selections.

\section{Inductive Predicate Learning}
\label{sec:predicate-learning}

\revised{In practice, analysts identify subgraphs of interest during analysis by either investigating which attribute features are inducing topology, e.g., a cluster or path, or examining how attribute distributions map to the overall topology.}
\revised{We capture this user analysis intent through selections.}
In this section, we demonstrate how our learning algorithm can synthesize these selections into predicates.

\begin{figure*}[t]
  \centering
  \includegraphics[width=0.9\linewidth]{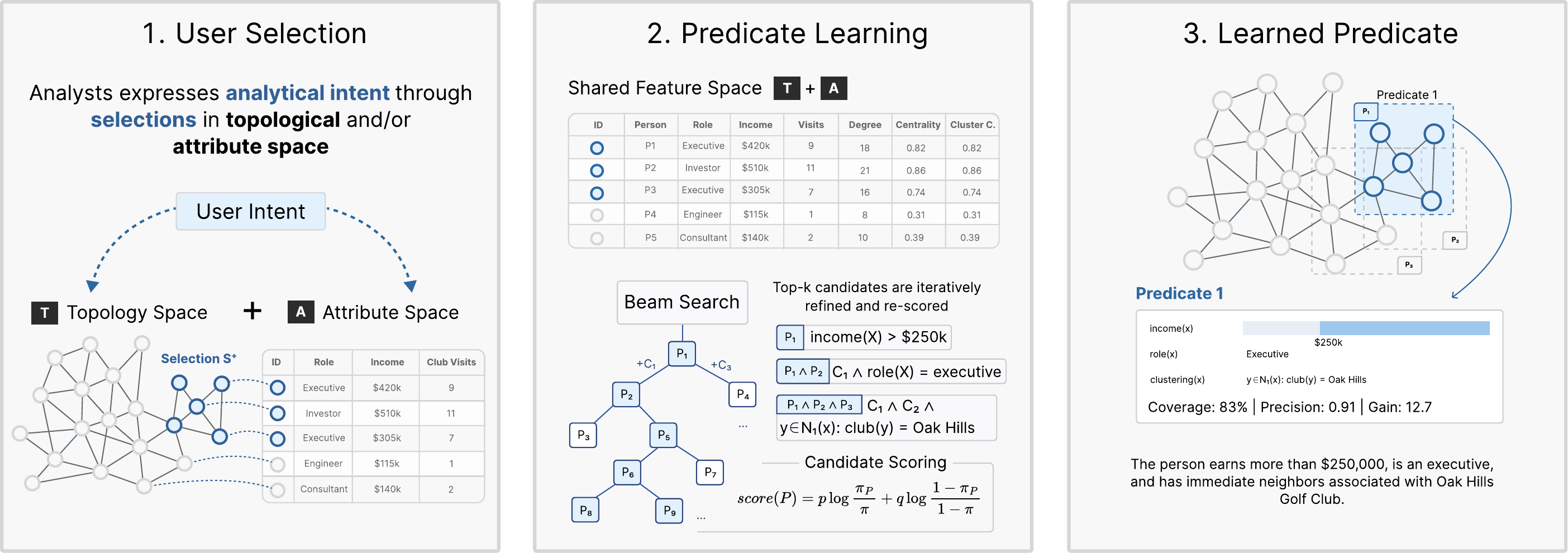}
  \vspace{-0.5em}
  \caption{\textbf{The predicate induction algorithm} takes a user-selected subgraph and translates it into a ranked set of predicate expressions that characterize structural and attribute-based patterns relative to the global graph. 
  }
  \label{fig:learning}
  \vspace{-2em}
\end{figure*}

\subsection{Learning Problem Description} \label{chap:learning_problem}

We formulate predicate learning as an inductive explanation problem. 
\revised{Given a multivariate graph $G = (V, E, A, R)$, a user selection $S^+ \subseteq V$ is obtained through interactions in topology space, attribute space, or both.}
\revised{Our objective is to identify predicate expressions describing what distinguishes many of the selected nodes from the rest. These predicates should capture structural and attribute-based patterns that occur more frequently within $S^+$ than expected.}
We model selection of each node $x \in V$ as an independent Bernoulli random variable indicating membership in $S^+$ with probability $\pi$.
Each candidate predicate $P(x)$ defines a subset \begin{equation}
    V_P = \{ x \in V \mid P(x) \}
    \label{eq:vp}
\end{equation}
\revised{of all nodes filtered by the predicate function $P(x)$.}
\revised{We can now relate $V_P$ to the selection $S^+$ by partitioning it into selected and unselected nodes.}
\revised{Using the cardinality operator, we define $p = |V_P \cap S^+|$ and $q = |V_P \setminus S^+|$, the numbers of correctly captured and overcaptured nodes by the predicate $P$. We can now distinguish two key metrics: the predicates precision $\pi_P = \frac{p}{p+q} = |V_P \cap S^+| / |V_p|$, and the global selection rate $\pi = |S^+| / |V|$, with $|V|$ nodes in total and $|S^+|$ nodes in the selection.}
\revised{We denote the unselected nodes as $S^- = V \setminus S^+$.}

A given selection may be characterized by multiple predicates.
We therefore identify a ranked set of predicates, each describing a subgraph that captures a higher proportion ($\pi_P$) than the global selection rate ($\pi$).
This avoids forcing a single global description and instead provides multiple complementary explanations of the selection.
Section~\ref{sec:scoring_model} describes the scoring model which determines predicate quality, and Section~\ref{sec:algo} presents the algorithm.

\subsection{Scoring Model} \label{sec:scoring_model}

\revised{Inspired by information retrieval ranking, we evaluate \emph{predicate quality} using the log-likelihood ratio under the Bernoulli selection model}:
\[
\textit{score}(P) =
p \log\left( \frac{\pi_P}{\pi} \right)
+
q \log\left( \frac{1-\pi_P}{1-\pi}\right)
\]
\revised{The first term captures \emph{enrichment}, the ratio between a predicate's recall rate and user's selection rate, that is $\pi_P / \pi$, then scaled by the number of user-selected nodes correctly captured by a predicate ($p$).
The second term penalizes a predicate's accidental coverage of unselected nodes.
Because our score scales with $p$ and $q$, a predicate must achieve both enrichment and sufficient \textit{support}, i.e., a high value for $|V_P|$, to rank highly.
We also report the fraction of selected nodes covered by a predicate as \emph{recall} ($\frac{|V_P \cap S^+|}{|S^+|}$), and the fraction of correctly captured nodes in the predicate as \emph{precision} ($\frac{|V_P \cap S^+|}{|V_P|}$).
To prevent overfitting to small subgraphs, we require each predicate captures at least $\tau=5$ nodes.
When the user selection is truly small in size, we use predicates that captures no more than $|S^+| + 2 $ nodes.
That is, we only consider candidate predicates $V_P$ with size $min(\tau, |S^+|+2) \leq |V_P| \leq max(\tau, |S^+|+2)$.
This ensures predicates have reasonable enrichment and support across varying selection sizes.}

\subsection{Predicate Induction Algorithm}
\label{sec:algo}

We construct predicates in two stages.
First, the induction algorithm identifies conjunctive clauses that describe selections.
Second, when a selection cannot be captured by conjunctive clauses, they are combined into disjunctions (see Section \ref{subsec:disjunctive}).

\subsubsection{Search Procedure}
In the first stage, each candidate predicate
$P(x)$ is a conjunction
of clauses drawn
from the attribute space, the topology space, and neighborhood-based predicates. 

\vspace{0.5em}\noindent\textbf{Generating Candidate Clauses.} \label{sec:thresholds}
All candidate clauses are first generated and scored individually.
For categorical attributes, one condition is generated per distinct observed value (e.g., $node\_type(x) = \text{``drug''}$, $node\_type(x) = \text{``protein''}$, etc.).
For numeric attributes, \revised{we derive candidate thresholds from the observed value distributions in the selected and unselected nodes.}
For example, predicates such as $degree(x) \geq t$ or $net\_production\_gwh(x) > t$ use thresholds $t$ obtained from these distributions (e.g., $t = 10$ or $t = 20{,}000$).
\revised{Let $z_1 < \cdots < z_d$ denote the distinct node attribute values in the graph. We generate thresholds at each midpoint $(z_i+z_{i+1})/2$ and additionally include the minimum, maximum, mean, median, and 25th and 75th percentiles of $S^+$.}
\revised{For each threshold, we compute lower and upper bound clauses using the log-likelihood ratio in Section~\ref{sec:scoring_model} and retain positive scores where the resulting clause covers at least $\tau$ nodes.}
This ensures that selected thresholds correspond to meaningful value ranges rather than arbitrary cut points.

\vspace{0.5em}\noindent\textbf{Generating Neighborhood Clauses.}
In addition to generating atomic clauses over topology and attributes, we generate clauses that capture patterns over node neighborhoods.
\revised{Neighborhood traversal is restricted to one-hop untyped neighborhoods and typed edge sequences of length at most two, which correspond to paths of up to two hops.}

For each base predicate $P_b(y)$, where $P_b(y)$ is any attribute or topology condition on a neighbor $y$, the learner generates three quantifier forms over the immediate one-hop neighborhood $N_1(x)$:\\
$\exists y \in N_1(x) : P_b(y),
\quad
\forall y \in N_1(x) : P_b(y),
\quad
|\{ y \in N_1(x) \mid P_b(y) \}| \geq j$.
The existential quantifier corresponds to $j = 1$ and the universal to $j = |N_1(x)|$; both are retained as explicit forms for readability.
\revised{We derive candidate values for $j$ from the distribution of neighbor counts in $S^+$ and $S^-$ following the same threshold selection procedure as for numeric attribute clauses.}
For graphs with typed edges, the same three forms are additionally generated over edge-type-constrained neighborhoods $N_\Sigma(x)$, where $\Sigma$ is an edge-type sequence of length at most two enumerated from the graph schema.
\revised{This limit is primarily a computational constraint: longer sequences yield exponentially many edge-type combinations, making exhaustive candidate generation intractable. One-hop untyped neighborhoods are sufficient for the presented use cases in Section \ref{chap:evaluation}. Extending to deeper untyped traversal is left for future work.}
\revised{Universal neighborhood filters are applied solely to nodes with neighbors, preventing automatic satisfaction.}

\vspace{0.5em}\noindent\textbf{Beam Search.}
\label{sec:search}
Predicate construction proceeds via beam search.
The beam maintains a fixed-size set of 5 conjunctive predicates, initialized from the top-scoring clause candidates.
\revised{We initialize the beam by greedily admitting the top-scoring clauses. A candidate is added only if its node-set Jaccard similarity with all previously admitted clauses remains below an empirical threshold of $\theta$=0.7.}
\revised{At each iteration, we extend and rescore every beam predicate with additional clauses.}
The beam retains the five highest-scoring extensions that satisfy the support threshold $\tau$ and achieve $\pi_P > \pi$. 
The search terminates when no refinement improves the score or when a maximum clause length of four is reached.
\revised{To mitigate order sensitivity introduced by the Jaccard similarity filter, candidates are sorted by score before seeding the beam.}

\subsubsection{Disjunctive Predicates} \label{subsec:disjunctive}

A single conjunctive predicate describes one subset of the selection at a time.
When the selected nodes contain multiple subgroups with distinct characteristics, a single predicate will rank highly for one group but leave the others unexplained.
In a second phase, we construct a disjunction of conjunctive predicates learned from Section \ref{sec:search} using a marginal-gain–based separate-and-conquer strategy\cite{furnkranz1999separate}.

Let $U$ denote the set of nodes covered by $P(x)$ added to the disjunction so far, initially empty.
\revised{We evaluate the predicate quality using the same support-weighted log-likelihood ratio defined in Section \ref{sec:scoring_model}, computed over the union of covered nodes.}
The global baseline selection rate remains fixed throughout learning.
For a candidate predicate $P$, its marginal contribution is the increase in log-likelihood ratio score when the nodes covered by $P$ are added to $U$.
\revised{At each iteration, beam search generates candidate clauses by maximizing the support-weighted log-likelihood ratio given the remaining uncovered, selected nodes.}
\revised{We add the predicate with the largest positive marginal gain to the disjunction.}
Learning terminates when no candidate yields further improvement, when a maximum of three conjunctive predicates has been reached, or when too few uncovered positives remain.
The final predicate is a disjunction of conjunctive predicates.
Because each conjunction in the disjunction must increase the union score, the predicate strictly improves the union score at every step.

Each candidate predicate is produced by the beam search and therefore already satisfies the minimum support threshold $\tau$ and $\pi_P > \pi$, as established in Section~\ref{sec:scoring_model}.
The one constraint specific to the disjunctive phase is that a predicate is only added if it yields a strictly positive marginal gain over the current union $U$, preventing re-description of nodes already covered by earlier predicates.
Together, these guarantees ensure that the learned disjunction consists of independently meaningful predicates rather than an arbitrary partition of $S^+$.

\section{The \toolname System} 
\label{chap:graphbridge}

\toolname is a visual analytics system for multivariate graph analysis.
It operationalizes the predicate language defined in Section~\ref{chap:predicates}, as well as the induction algorithm introduced in Section~\ref{sec:predicate-learning}. 
A typical workflow in \toolname combines visual exploration across topology and attribute spaces with predicate construction and induction.
Analysts make selections in either space, derive predicate descriptions, and iteratively refine both the selection and the corresponding predicate.



\subsection{System Overview} 



The interface consists of four coordinated panels (see Figure \ref{fig:system}). 
The \emph{topology panel} (A.1) shows graph topology through a node-link visualization with multiple layout algorithms and an aggregated schema view of connections between node types, supporting analysis of connectivity patterns and neighborhood structure.
The \emph{attribute panel} (A.2) presents distributions of node attributes, enabling characterization of selections and identification of relevant attribute ranges.
\revised{The attribute panel orders node-type sections by total node count and attributes within each section alphabetically.
Within distributions, categorical values are ordered by selected-node count, falling back to overall count when no selection is active, so that the most relevant values appear first.}

The \emph{predicate panel} (B) features the predicate builder (B.1) and induction results (B.2). 
The visual predicate builder features a no-code interface for manually constructing expressions in the predicate language defined in Section \ref{chap:predicates}.
See Section \ref{chap:predicate_builder} 
for a description of the language constructs available through the interface.
The predicate induction panel (B.2) presents predicates learned from user selections, ranked by their enrichment values, and accompanied by coverage and support statistics (see Section \ref{chap:learning_problem}).
These representations allow analysts to inspect, compare, and refine candidate predicates.

\subsection{Constructing Expressions in the Predicate Builder}
\label{chap:predicate_builder}


To support predicate construction without requiring users to write or debug formal expressions, \toolname provides a visual predicate builder.
The builder enables the specification of predicates over attributes, topology, and neighborhoods.
All elements of the predicate language (see Section~\ref{chap:predicates}) are exposed through drag-and-drop operations, menus, and parameter controls, ensuring that all constructed expressions are valid.



\vspace{0.5em}\noindent\textbf{Predicate Clauses.}
In the predicate builder, atomic clauses are represented as \emph{pills}, each corresponding to a single condition over the loaded graph.
When an analyst observes a pattern in the topology or attribute panel, a pill can be created directly from that panel, grounding the clause in a concrete data observation.
Alternatively, pills can be added through the builder's menu when the analyst already has a specific condition in mind.
The builder distinguishes green attribute pills from blue topology pills, mapping to the attribute and topology spaces respectively.
Categorical values are selected from a dropdown populated from the graph, and numeric thresholds are adjusted using a slider anchored to the value distribution.

\vspace{0.5em}\noindent\textbf{Neighborhood Predicates.}
A neighborhood block wraps an inner predicate with a quantifier over a node's neighbors.
The analyst selects the quantifier ($\exists$, $\forall$, or count $\geq k$) from a dropdown, specifies the neighborhood scope from a menu of available edge types drawn from the graph schema, and places a pill inside the block as the inner condition.
This produces predicates such as ``at least one neighbor of type \textit{protein} has degree above 4'' without requiring the analyst to write a quantified expression by hand.

\vspace{0.5em}\noindent\textbf{Combining Predicates.}
Pills are placed on a canvas and linked with AND and OR connectors to form conjunctions and disjunctions.
Any set of pills can be grouped using parentheses, and groups may be nested to arbitrary depth.
An analyst can construct conditions such as ``belongs to community 3 AND (degree $\geq$ 50 OR has at least one neighbor with degree $\geq$ 200)'' through direct manipulation.


\vspace{0.5em}
Upon execution, constructed expression by the user is evaluated over the graph.
The backend maintains an internal tree representation of pills, groups, and neighborhood blocks. 
This structure is serialized into the formal predicate syntax and parsed into a typed abstract syntax tree.
Evaluation of the expression proceeds recursively. 
Topological metrics are computed on first access and cached for the lifetime of the loaded graph.
Conjunction evaluation stops at the first false operand and disjunction evaluation stops at the first true operand.
For graphs containing several thousand nodes, predicate evaluation completes in under 100 milliseconds on a standard workstation, enabling interactive refinement in \toolname.
\revised{Predicate induction takes between 1 and 3 seconds depending on graph size, run on the same hardware.}


\subsection{Automated Predicate Induction Through Interactions} \label{sec:system-induction-interactions}

Analysts frequently encounter patterns they recognize before they can describe them.
A structurally central cluster, a bridge between two communities, or an unusual attribute distribution may be visible in the topology or attribute panel.
\toolname externalizes such observations through predicate induction, linking each selection made by the analyst to a ranked list of descriptive expressions.
Each result in the induction panel displays the predicate expression alongside its enrichment, coverage, and support statistics, as defined in Section~\ref{chap:learning_problem}.
Any result can be inserted into the predicate builder as an editable expression, turning a automatically generated predicates into a starting point for refinement.
\toolname supports three interactions that determine how the selection is contrasted during induction.


\vspace{0.5em}\noindent\textbf{Selection.}
The analyst selects any subgraph of interest by clicking, lassoing, or filtering in either the topology or attribute panel, with all other nodes serving as background.
The system induces predicates drawing on conditions from both the topology and attribute spaces (Section~\ref{sec:predicate-learning}), surfacing cross-space characterizations that span both spaces jointly.
This is the primary mode of exploration, turning any visual observation into a formal description.

\vspace{0.5em}\noindent\textbf{Comparison.}
Some analytical questions concern not what defines a group globally, but what distinguishes it from a specific other group.
The analyst designates a second subgraph ($S^+_2$) as a contrastive baseline, and the system learns predicates that are enriched in the primary selection relative to that baseline rather than the graph as a whole (formally, replacing $V$ in Equation~\ref{eq:vp} with $S^+_2$).
Executing such a predicate retrieves all nodes in the graph that share the distinguishing properties, whether or not they appeared in either original selection.
This mode is suited to comparing two groups that share many global properties but differ in targeted ways, where a global enrichment signal would obscure the relevant distinction.

\vspace{0.5em}\noindent\textbf{Path Selections.}
When exactly two nodes are selected, \toolname offers a path-finding dialog to expand the selection.
The analyst retrieves all shortest paths or enumerates paths within a specified hop range.
The intermediate nodes along those paths form the selection.
The system characterizes these intermediaries through induction, revealing what structural and functional properties they share.
This answers questions such as which proteins mediate a mechanistic path between a drug and a disease target, and what makes them suitable as intermediaries.

\section{Use-Cases}
\label{chap:evaluation}

We evaluate \toolname through three use cases in energy systems, cybersecurity, and drug repurposing.
\revised{For each case, a domain expert provided a dataset and described representative analytical scenarios during an hour-long open-ended session.
These experts are further described in Section~\ref{chap:expert_feedback}.}
We report statistics following the definitions in Section~\ref{chap:learning_problem}.
\revised{The use-case analyses are descriptive rather than predictive. 
We therefore report both precision and coverage: high precision excludes unselected nodes, whereas high coverage retains more of the analyst's selection.
Uncovered selected nodes can be understood as false negatives relative to that selection.}

\subsection{Legacy Fleet Risk in Wind Transmission Networks}
\label{sec:energy}

European electricity grid operator \href{https://www.tennet.eu/}{TenneT} manages a transmission network powered by thousands of wind turbines\cite{tennet_nh_energy}.
A common question for grid planning is ``what types of turbines connect to the most critical substations in the network''.
\revised{We demonstrate how \toolname supports this analysis through a workflow in which a topological observation leads to a formal characterization of a legacy group of turbines whose age, low output, and concentration at backbone substations could create disproportionate grid consequences.}

\vspace{0.5em}\noindent\textbf{Dataset.}
We use a graph dataset provided by TenneT\cite{tennet_nh_energy}, which contains 5,279 nodes and 5,168 edges.
Turbine nodes have attributes like net production, capacity, turbine model, manufacturer, and commissioning year.
Substation nodes encode voltage level and operational status.
Edges encode directed generation flows and high-voltage transmission connections between substations.

\vspace{0.5em}\noindent\textbf{Analysis.}
\revised{The analyst inspects the topology view and identifies ZWO150, an onshore substation in Zeewolde, as the substation connected to the largest number of wind turbines (see Figure~\ref{fig:system}A).}
Inspecting ZWO150 shows 222 wind turbines feeding into it average only around 4,500~MWh per connection, compared to 20,500~MWh per connection at other substations, such as sub\_HNA220 (an offshore wind park).
In other words, the wind turbines connected to substation ZWO150 contain low-efficiency turbines.
Using the visual predicate builder, the analyst selects 437 turbines with net production below 7,000~MWh that feed into substations with degree above 80 (see Figure~\ref{fig:system}B).
This results in 437 turbines feeding into major substations.
The analyst observes that the Zeewolde fleet constitutes the majority of this selection and continues the investigation using \toolname to investigate properties of this fleet.

\vspace{0.5em}\noindent\textbf{Induction.}
Applying predicate learning over the 437 selected turbines yields five predicates.
The top-ranked predicate is broad, matching 422 of the selected turbines but also 665 outside the selection, and yields 38.8\% precision.
The second-ranked predicate achieves perfect precision:
\begin{equation*}
\begin{aligned}
\texttt{manufacturer}(x) &= \text{``Neg Micon''} \\
\wedge\;&\exists y \in N_{(\texttt{feeds\_into})}(x) : \texttt{pagerank}(y) > 0.0074
\end{aligned}
\end{equation*}
\vspace{-2em}
\begin{flushright}
\textnormal{[p=115, q=0, precision 100\%, enrichment 12.1$\times$, coverage 26.3\%]}
\end{flushright}
The manufacturer clause alone yields 61.5\% precision, with 72 of the 187 Neg Micon turbines appearing as false positives due to connections to smaller substations outside the transmission backbone.
\brian{Of the 187 Neg Micon turbines captured in the manufacturer clause, 72 are connected to smaller substations outside the transmission backbone characterized by the 437 selected turbines (precision=61.5\%).}
Adding the PageRank clause on the connected substation removes these false positives, resulting in perfect precision.
\brian{Including the PageRank clause reduces the number of turbines to 115. Of the remaining turbines, all were found in the 437 selected turbines (precision=100\%).}
\revised{The value $0.0074$ is a dataset-specific learned boundary rather than a universal grid threshold.
In this graph, it separates globally central backbone substations from smaller peripheral substations.
The predicate therefore reveals that Neg Micon turbines become distinctive specifically when they feed into transmission hubs with high network-wide importance.}

The fourth-ranked predicate identifies the same turbines from the turbine model rather than the manufacturer.
\begin{equation*}
\begin{aligned}
\texttt{turbine\_type}(x) &= \text{``V52-850~kW''} \\
\wedge\;&\exists y \in N_{\texttt{(feeds\_into)}}(x) : \texttt{degree}(y) > 84
\end{aligned}
\end{equation*}
\vspace{-2em}
\begin{flushright}
\textnormal{[p=89, q=0, precision 100\%, enrichment 12.1$\times$, coverage 20.4\%]}
\end{flushright}
\revised{The turbine-type clause alone yields 50.6\% precision because 88 V52-850~kW units at secondary hubs are false positives.}
\brian{Of the XXX V52-850~kW captured by the turbine-type clause, 88 are connected to secondary hubs (precision=50.6\%).}
\revised{Adding the substation-degree clause removes them, producing perfect precision.}
\brian{Including the substation-degree clause reduces the number of turbines to 89, all of which were included in the selected turbines making up the transmission backbone (precision=100\%).}
\revised{The manufacturer and turbine-type clauses identify complementary subsets of the same legacy fleet when combined with a structural clause.}
\revised{These zero-false-positive predicates are highly specific but narrow: they cover 26.3\% and 20.4\% of the 437 selected turbines, respectively. By contrast, the top-ranked predicate recovers most of the selection but includes more false positives.}
\brian{These predicates are characterized by 100\% precision but low coverage, capturing 26.3\% and 20.4\% of the 437 selected turbines. By contrast, the top-ranked predicate captures most of the selection (XXX\%) at the cost of lower precision (XXX\%)}

\vspace{0.5em}\noindent\textbf{Takeaway.}
\revised{Neg Micon, a major Danish manufacturer in the 1990s and early 2000s, represents one of the dataset's oldest turbine segments.}
Neither age nor maintenance status appears as a node attribute, yet the induction algorithm recovers this legacy group from manufacturer identity and substation structure alone.
\revised{Individually these turbines have low output, but their concentration at backbone substations combined with correlated degradation risk across the same generation means that simultaneous failures could create a high-consequence grid event.
By combining topology and attribute information, the induction algorithm identifies Neg Micon and V52-850~kW turbines at backbone substations as a concrete risk pattern: legacy, low-output turbine families concentrated at structurally important aggregation points.}


\subsection{Characterizing Threat Actors in Cybersecurity}
\label{sec:cybersecurity}

APT41\cite{mitre_apt41} and Volt Typhoon\cite{mitre_volt_typhoon} are both Chinese state-sponsored cybersecurity threat actors.
Although they operate within the same geopolitical context, their techniques and tactics differ in ways that affect detection strategies.
We demonstrate how \toolname determines whether their technique repertoires exhibit differences that require reasoning across both the attribute space and the topology space.

\vspace{0.5em}\noindent\textbf{Dataset.}
We use the MITRE ATT\&CK graph\cite{mitre_attack_stix_data}, which contains 1,742 nodes and 19,215 edges.
Node types include threat actors, techniques, tools, and mitigation strategies.
Nodes have multiple attributes. For example, nodes of type \emph{technique} encode attributes such as a name, the platform it targets, and the associated tactics.
Edges encode relationships between actors, malware, techniques, and more.

\vspace{0.5em}\noindent\textbf{Analysis.}
The analyst locates both threat actors in the topology view and expands the selection to their directly connected nodes of type \emph{technique}.
The attribute view shows how the selected techniques distribute across tactics, the high-level adversarial goals each technique is associated with, such as privilege escalation and defense evasion. 
This reveals 82 techniques associated with APT41 and 81 with Volt Typhoon, with 35 techniques shared between the two actors.
The analyst focuses on the techniques unique to each actor.
This yields 47 APT41 and 46 Volt Typhoon techniques.

\begin{figure}[t]
  \centering
  \includegraphics[width=\linewidth]{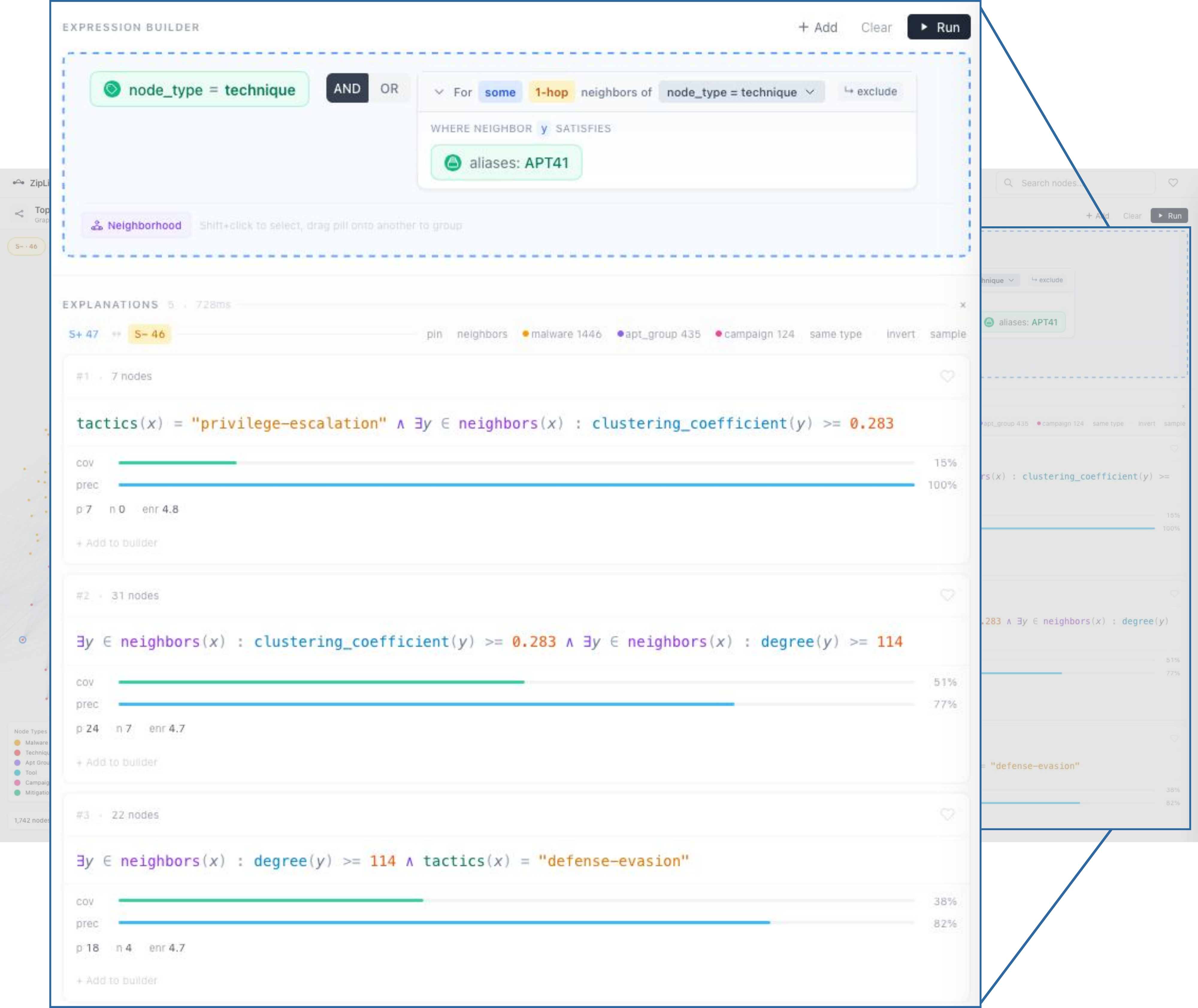}
  \caption{Top-ranked predicates distinguishing APT41 from Volt Typhoon combine tactic-based attributes with structural neighborhood filters.}
  \label{fig:cybersec_usecase}
  \vspace{-2em}
\end{figure}

\vspace{0.5em}\noindent\textbf{Induction.}
Contrasting the 47 APT41 techniques with the 46 Volt Typhoon techniques using \toolname yields five predicates. 
\revised{Figure~\ref{fig:cybersec_usecase} shows the first three.}
The top-ranked predicate achieves perfect precision:
\begin{equation*}
\begin{aligned}
\texttt{tactics}(x) &= \text{``privilege-escalation''} \\
\wedge\;&\exists y \in N_1(x) : \texttt{clustering\_coefficient}(y) \geq 0.283
\end{aligned}
\end{equation*}
\vspace{-2em}
\begin{flushright}
\textnormal{[p=7, q=0, precision 100\%, coverage 14.9\%]}
\end{flushright}
The tactic clause alone yields 83.3\% precision, as 2 of the 12 matched privilege-escalation techniques belong to Volt Typhoon.
Adding the neighborhood clustering clause eliminates these 2 false positives by isolating APT41 privilege-escalation techniques embedded in dense malware clusters.
\brian{Of the 12 privilege-escalation techniques captured by the tactics clause, 10 were associated with APT41 while the remaining 2 were associated with Volt Typhoon (precision=83.3\%). Including the neighborhood clustering clause further isolates privilege-escalation techniques embedded in dense malware clusters. Of these 7 techniques, all were associated with APT41 (precision=100\%).}
\revised{Under \toolname's metric, a clustering coefficient of $0.283$ means that at least 28.3\% of possible links among the neighboring node's neighbors are present.
Here, the threshold separates sparse from locally dense malware or tool neighborhoods, revealing a connected operational context.}
The third predicate covers a broader pattern across defense-evasion techniques:
\begin{equation*}
\begin{aligned}
\texttt{tactics}(x) &= \text{``defense-evasion''} \\
\wedge\;&\exists y \in N_1(x) : \texttt{degree}(y) \geq 114
\end{aligned}
\end{equation*}
\vspace{-2em}
\begin{flushright}
\textnormal{[p=18, q=4, precision 81.8\%, coverage 38.3\%]}
\end{flushright}
Defense evasion alone yields 72.0\% precision, as 7 of the 25 matched techniques belong to Volt Typhoon.
\brian{The tactics clause alone yields 72.0\% precision, as 7 of the 25 matched defense-evasion techniques were associated with Volt Typhoon.}
APT41 defense-evasion techniques connect to malware families and tools with high degree, reflecting broad adoption across the threat-actor ecosystem, while Volt Typhoon’s equivalents fall below the degree threshold of 114.
\revised{More concretely, degree(y)$\geq114$ requires a neighboring malware or tool node linked to at least 114 entities in the ATT\&CK graph. The condition therefore operationalizes broad reuse or co-reference across the threat-intelligence ecosystem.}
That two independent tactic categories converge on the same structural pattern confirms that the distinction reflects an operational difference rather than an artifact of any single tactic.
\revised{The two predicates also illustrate the precision-coverage trade-off. The first has 100\% precision but covers only 14.9\% of the APT41-only techniques. The second admits four Volt Typhoon techniques but covers 38.3\% of the APT41 selection. The broader predicate may therefore provide a more representative characterization of the actor's repertoire even though its precision is lower.}

\vspace{0.5em}\noindent\textbf{Takeaway.}
The predicates learned by \toolname reveal a distinction between the two actors.
APT41's techniques, across both privilege escalation and defense evasion, are embedded in densely interconnected regions of the graph, reflecting broad co-referencing across malware families and tools.
Volt Typhoon's techniques occupy structurally peripheral positions, consistent with a more targeted operational profile oriented towards exploitation and credential abuse.
The separation emerges only when tactic attributes and structural conditions are combined, confirming that distinguishing these actors requires reasoning across both the attribute and topology spaces.
For detection engineering, this distinction implies different monitoring strategies.
APT41 activity can be detected through existing malware signatures and threat intelligence feeds.
Volt Typhoon activity requires behavioural monitoring focused on credential misuse and exploitation events.
\revised{The learned thresholds make this distinction operational: clustering identifies locally cohesive malware contexts, whereas degree identifies tools or malware reused across many graph entities. 
Their contribution is therefore explanatory, not merely an increase in precision.}

\subsection{Drug Repurposing via Molecular Bridges}
\label{sec:drug_repurposing}

Drug repurposing requires identifying molecular targets that connect disease contexts through shared pharmacological mechanisms.
In many cases, off-label effects are observed and known, but the underlying structural mechanism remains unclear.
We use \toolname to understand why a retinoid approved for a blood cancer also affects a T-cell lymphoma of the skin.

\vspace{0.5em}\noindent\textbf{Dataset.}
We use a subset of the PrimeKG graph\cite{chandak2023building}, which contains 1,538 nodes and 12,570 edges.
Node types include drug targets, diseases, and genes and proteins.
Nodes have multiple attributes.
For example, nodes of type gene and protein have attributes such as molecular function, biological process, and cellular component.
Edges encode relations including drug-protein binding, off-label use associations, disease-gene links, and protein-protein interactions.
The subset used in this analysis contains the 100 highest-prevalence diseases with their associated nodes, concentrating the graph around relevant relationships.

\vspace{0.5em}\noindent\textbf{Analysis.}
The analyst selects the \emph{Tretinoin} node in the attribute view and opens the path finder (see Section \ref{sec:system-induction-interactions}) to \emph{primary cutaneous T-cell lymphoma} as the target endpoint.
\revised{Setting the path length between 2 and 4 bypasses the direct off-label use edge while retaining up to two molecular intermediaries, yielding 108 selected nodes across 500 paths.}
The analyst filters this selection to retain only 23 drug targets directly connected to Tretinoin, 14 disease genes associated with CTCL, and 11 that appear in both groups.

\vspace{0.5em}\noindent\textbf{Induction.}
The analyst first runs the predicate induction algorithm over the 23 drug target nodes, returning:
\begin{equation*}
\begin{aligned}
\texttt{molecular\_functions}(x) &= \text{``iron ion binding''} \\
\wedge\;&\texttt{node\_type}(x) = \text{``drug\_target''} \\
\wedge\;&|\{\, y \in N_{\texttt{(drug-protein.indication)}}(x): \\
 &\texttt{degree}(y) \geq 63.5 \,\}| \geq 2
\end{aligned}
\end{equation*}
\vspace{-2em}
\begin{flushright}
\textnormal{[p=13, q=4, precision 76.5\%, enrichment 51.1$\times$, coverage 56.5\%]}
\end{flushright}
The iron ion binding condition alone yields 54.2\% precision, as 11 of the 24 matching nodes in the graph fall outside the drug target selection.
Iron ion binding is the molecular signature shared by all cytochrome P450 enzymes, the proteins responsible for breaking Tretinoin down in the body.
All 13 matched genes carry a \emph{retinoic acid metabolic process} annotation, confirming them as the enzymes that process the drug.
The ten drug targets not covered by this predicate carry \emph{cellular response to retinoic acid} annotations.
These are the receptors Tretinoin directly activates, a distinct role the predicate separates automatically.

Learning over the 14 disease gene nodes returns:
\begin{equation*}
\begin{aligned}
&|\{\, y \in N_{\texttt{(disease-protein.off-label)}}(x): \\
& 
\texttt{louvain\_community}(y) = \text{``cluster\_6''} \,\}| \geq 2 \\
\wedge\;&\exists z \in N_1(x) : \texttt{pagerank}(z) \geq 0.0086
\end{aligned}
\end{equation*}
\vspace{-2em}
\begin{flushright}
\textnormal{[p=13, q=3, precision 81.2\%, enrichment 89.3$\times$, coverage 92.9\%]}
\end{flushright}
The neighbourhood condition alone yields 50.0\% precision, as 13 of the 26 matching nodes fall outside the disease gene selection.
The analyst inspects the cluster\_6 nodes in the topology view and finds that they are all drugs used to treat cutaneous T-cell disease.
The predicate recovers 13 of 14 known CTCL cancer genes by identifying genes that sit next to those drugs in the graph, without using any cancer-specific biological annotation.
CREBBP and SMARCA4 are direct binding partners of the retinoic acid receptor, a connection the system finds through graph structure alone.
\revised{This predicate deliberately favors coverage over perfect precision: it retains 13 of the 14 disease genes (92.9\% coverage) while admitting three false positives. For hypothesis generation, this trade-off is useful because a small number of additional candidates can be inspected manually, whereas omitting a plausible molecular bridge may be more costly.}
\revised{Applying induction to the 11 bridge gene nodes yields a predicate based solely on the k-core topological metric. This indicates that the bridge genes occupy a cohesive core of the graph because the predicate provides no attribute-level characterization.}

\vspace{0.5em}\noindent\textbf{Takeaway.}
The predicates together trace the molecular mechanism behind Tretinoin's off-label activity in CTCL.
The iron ion binding predicate identifies the CYP enzyme cluster responsible for metabolising the drug.
The cluster\_6 neighbourhood predicate recovers CTCL disease genes through their proximity to CTCL-specific drugs in the graph, without any cancer-specific annotation.
This chain of evidence is not recoverable from the off-label edge or from path inspection alone.
For a drug repurposing analyst, the predicates provide a formal and reusable molecular hypothesis applicable to other retinoid-disease pairs.

\begin{figure}[t]
  \centering
  \includegraphics[width=\linewidth]{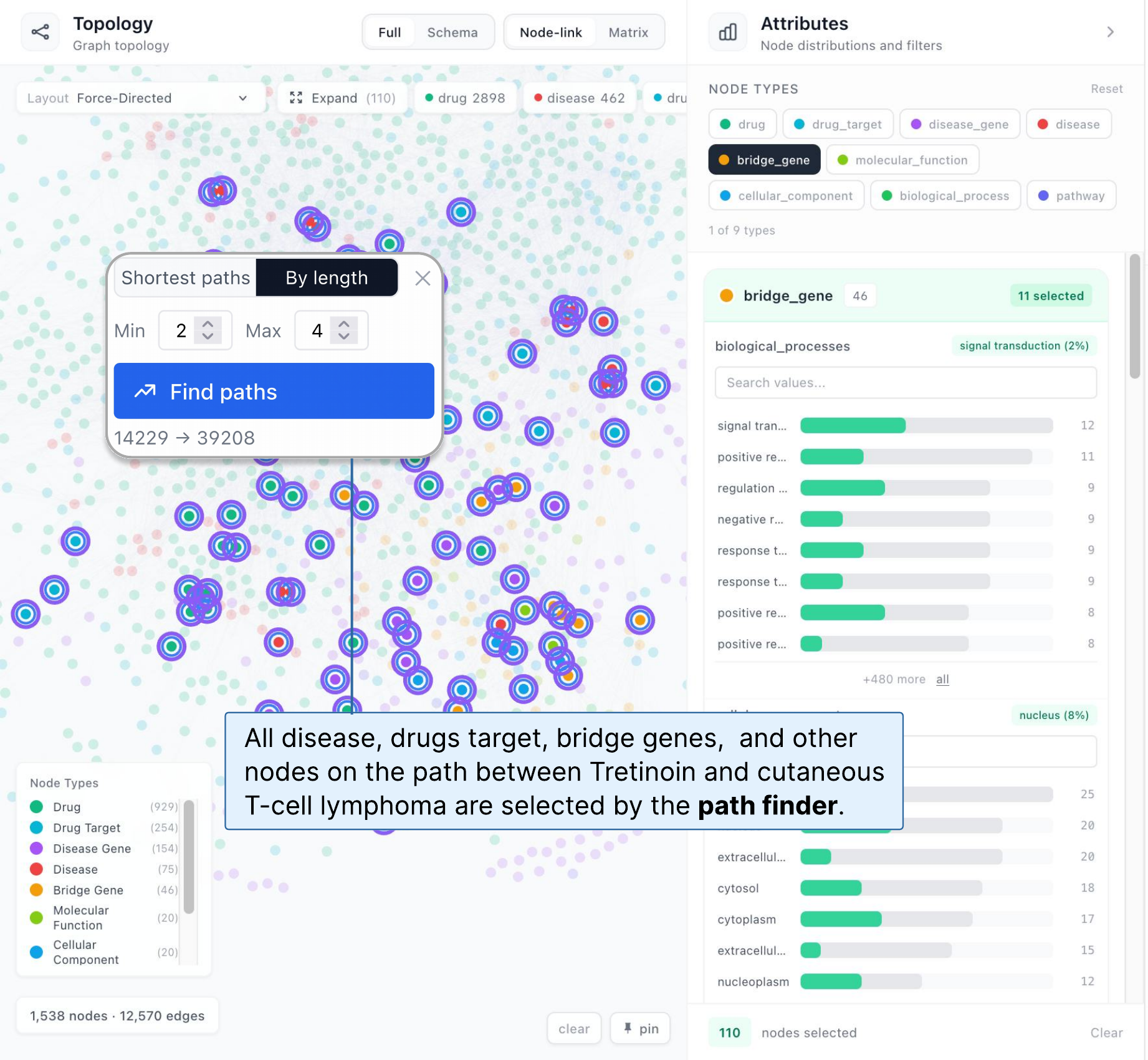}
  \caption{
  Path finder selects drug targets, disease genes, and shared intermediaries that link Tretinoin to CTCL.
}
  \label{fig:drugrepurposing_usecase}
  \vspace{-2em}
\end{figure}

\section{Expert Feedback} \label{chap:expert_feedback}

\revised{We conducted preliminary expert walk-throughs with the three domain experts introduced in Section \ref{chap:evaluation}.
Expert E1 (grid planning engineer) is affiliated with [anonymized] National Laboratory and has more than six years of experience in energy infrastructure analysis.
Expert E2 (cybersecurity researcher) is affiliated with [anonymized] National Laboratory and has three years of experience in cybersecurity analysis.
Expert E3 (immunology researcher) is affiliated with [anonymized] Medical School and has four years of experience in immunology.}
\revised{None of the experts had prior experience with \toolname.
All experts were familiar with graph visualization tools such as Gephi or Cytoscape but had no prior exposure to predicate-based query systems.
Each session consisted of a guided walk-through of a representative use case, including a demonstration of the system and its predicate induction capabilities.}
\revised{Experts were asked to assess (1) whether the learned predicates aligned with their domain knowledge, (2) whether \toolname would fit their existing workflows, and (3) whether the interface was understandable.
We also solicited feedback on limitations and potential improvements.
Each session lasted approximately 45 minutes.
No compensation was provided.}

Across domains, experts found \toolname useful and considered the results of the induction algorithm meaningful and consistent with domain knowledge.
The learned predicates were seen as applicable to real analysis tasks, and the system was perceived as providing capabilities that are not readily available in existing tools.
E2 emphasized that \textit{“there aren’t many systems that prioritize flexibility, control, and filtering”} for their domain.
E1 described the ability to derive alternative predicate explanations as \textit{“super useful and interesting”}.
The separation into topology and attribute spaces was consistently understood and considered effective, with their integration supporting analysis across both perspectives.

\vspace{0.5em}\noindent\textbf{Interpretability.}
The predicate builder was considered an important component of the workflow and was viewed as intuitive to use.
E2 noted that the workflow \textit{“follows how I would explore the data myself”}.
In contrast, the results of the induction algorithm required more effort to interpret and were perceived as less immediately trustworthy.
\revised{In particular, translating conjunctions of multiple predicates (especially those combining structural and attribute conditions) into domain meaning required considerable effort.}
Experts also noted that the relationship between learned predicate statistics (e.g., precision, coverage) and the underlying selection was not always clear at first, although it became useful once understood.
E1 noted that \textit{“interpreting multiple explanations and comparing them may require additional visualization techniques”}.
In addition, users may expect that predicates constructed manually are reproduced exactly by the induction process, indicating that the distinction between constructed and learned predicates requires clearer communication.

\vspace{0.5em}\noindent\textbf{Topology Features.}
While predicates over attributes were generally easy to interpret, topology-based components required more domain context.
E3 noted that general-purpose topological metrics, beyond simple measures such as degree, \textit{"are difficult to interpret when presented as part of learned predicates"}.
In contrast, degree was considered meaningful in targeted expressions within the predicate builder, for example when retrieving nodes within a specific degree range.
At the same time, E1 expressed interest in extending the system with \textit{"additional topology metrics tailored to domain-specific analysis tasks"}.
Together, these observations suggest that the interpretability of topology depends on both the form of presentation and the availability of domain-relevant structural measures.
Predicate expressions involving neighborhood conditions were also more difficult to interpret.
In addition, communicating results based on topological metrics to decision makers was identified as a challenge. 
As E1 and E3 observed, \textit{“the metrics may live in the graph space rather than what they mean in the application domain”}.
These observations indicate a need for improved representations of learned predicates, for example through visual summaries or natural language explanations.
At the same time, E1 and E2 emphasized the value of linking both spaces, describing the ability to \textit{“link attribute subsets with the network”} as \textit{“super powerful”}.

\vspace{0.5em}\noindent\textbf{Areas of Improvement.}
Experts identified several opportunities to improve the system.
Domain-specific extensions were suggested, including support for temporal analysis and mechanisms to capture evolving data, as E2 noted the need to \textit{“highlight the dynamic nature”} of the domain.
The relevance of attributes was noted to depend on the analysis task, motivating mechanisms to filter or prioritize attributes within the interface. 
Further improvements include enhanced visualization controls, clearer documentation, and support for natural language input to facilitate predicate construction.

\section{Discussion, Limitations, and Future Work}

We presented \toolname, a system for multivariate graph analysis that integrates reasoning across topology and attribute spaces. 
The system operationalizes a predicate language and an induction algorithm that together enable analysts to express, derive, and refine patterns that combine structural and attribute-based properties.
By bridging these two spaces within a single workflow, \toolname supports iterative analysis in which visual exploration and formal specification inform each other.

Across the use-cases in Section \ref{chap:evaluation}, domain experts confirmed that the derived patterns are meaningful and consistent with domain knowledge.
The system was perceived as useful and provided capabilities that are not readily available in existing tools.
In particular, the predicates learned by the induction algorithm were considered valuable and applicable within the respective domains, supporting the analysis.

At the same time, domain experts express that the attributes are generally intuitive to interpret, as they correspond directly to domain concepts, whereas topological metrics, such as centrality, cluster-coefficients, are less readily translated into real-world meaning.
This gap will become more pronounced when communicating results from \toolname to decision makers who are not familiar with topology metrics. 
Even for technically skilled users, predicate expressions that combine structural and attribute-based conditions can be difficult to interpret. 
This highlights an opportunity to improve interpretability through complementary representations, such as visual summaries of learned predicates or natural language explanations.
Additionally, enabling natural language input for predicate construction may further lower the barrier to expressing complex expressions.

The expert feedback indicates that \toolname is a general-purpose system that is broadly useful across domains.
Experts recognized its potential and identified opportunities to tailor the system to domain-specific needs.
Certain domains require specialized reasoning capabilities, such as temporal analysis and dependency modeling.
These requirements are not limitations of \toolname, but reflect opportunities for extension, as the underlying formalism and induction algorithm can accommodate this or be extended.
Experts also noted that the system inherits challenges common to graph visualization. 
In particular, the interpretation of topology depends on the chosen layout, which may influence analysis.
This is not specific to \toolname, but reflects a broader limitation of graph visualization.

Finally, the relevance of attributes depends on the specific analysis task.
Not all attributes are equally meaningful for predicate learning, and some, such as spatial coordinates, may not contribute to certain analytical goals at all.
This suggests an opportunity to allow users to guide the learning algorithm by weighting or prioritizing attributes based on domain knowledge and task requirements.
Incorporating such mechanisms is left for future work.

\section{Conclusion}

Analysis of multivariate graphs requires reasoning jointly over two representational spaces: topology and attributes. 
\revised{Existing systems often treat topology and attribute exploration as complementary but distinct activities.}
\toolname addresses this by integrating predicate construction and automated predicate induction for multivariate graph analysis. 
The system features a predicate builder for expressing filters over attributes, structural properties, and neighborhood relations. 
An induction algorithm learns logical predicates from analyst selections.
By combining interactive predicate specification with induction, \toolname enables analysts to move from visual exploration to formal reasoning about topological-attribute patterns.

\bibliographystyle{abbrv-doi-hyperref}

\bibliography{template}


\end{document}